\documentclass[11pt]{article}
\pdfoutput=1  % for arXiv
\usepackage[utf8]{inputenc}
\usepackage[english]{babel}
\usepackage{amsmath, bm}
\usepackage{graphicx}
\usepackage{authblk}
\usepackage{hyperref}
\usepackage{cite}
\usepackage{geometry}

\newgeometry{left=2.5cm,right=2.5cm, top=2.5cm, bottom=3.5cm}   % set the margins

\begin{document}
\title{Learning neural network potentials from experimental data via Differentiable Trajectory Reweighting}
\author[1]{Stephan Thaler\footnote{stephan.thaler@tum.de} }
\author[1,2]{Julija Zavadlav\footnote{julija.zavadlav@tum.de} }
\affil[1]{Professorship of Multiscale Modeling of Fluid Materials, \linebreak
TUM School of Engineering and Design, Technical University of Munich, Germany}
\affil[2]{Munich Data Science Institute, Technical University of Munich, Germany}
\date{}  % date does not appear
\renewcommand\Affilfont{\itshape\normalsize}

\maketitle

\begin{abstract}
In molecular dynamics (MD), neural network (NN) potentials trained bottom-up on quantum mechanical data have seen tremendous success recently.
Top-down approaches that learn NN potentials directly from experimental data have received less attention, typically facing numerical and computational challenges when backpropagating through MD simulations. 
We present the Differentiable Trajectory Reweighting (DiffTRe) method, which bypasses differentiation through the MD simulation for time-independent observables. Leveraging thermodynamic perturbation theory, we avoid exploding gradients and achieve around 2 orders of magnitude speed-up in gradient computation for top-down learning.
We show effectiveness of DiffTRe in learning NN potentials for an atomistic model of diamond and a coarse-grained model of water based on diverse experimental observables including thermodynamic, structural and mechanical properties. Importantly, DiffTRe also generalizes bottom-up structural coarse-graining methods such as iterative Boltzmann inversion to arbitrary potentials.
The presented method constitutes an important milestone towards enriching NN potentials with experimental data, particularly when accurate bottom-up data is unavailable.

\end{abstract}

\section{Introduction}
Molecular modeling has become a cornerstone of many disciplines, including computational chemistry, soft matter physics, and material science.
However, simulation quality critically depends on the employed potential energy model that defines particle interactions.
There are two distinct approaches for model parametrization \cite{Frohlking2020, Noid2013}: 
Bottom-up approaches aim at matching data from high fidelity simulations, providing labeled data of atomistic configurations with corresponding target outputs. Labeled data allows straightforward differentiation for gradient-based optimization, at the expense of inherently limiting model accuracy to the quality imposed by the underlying data-generating simulation.
On the other hand, top-down approaches optimize the potential energy model such that simulations match experimental data. From experiments, however, labeled data on the atomistic scale are not available. Experimental observables are linked only indirectly to the potential model via an expensive molecular mechanics simulation, complicating optimization.

A class of potentials with tremendous success in recent years are neural network (NN) potentials due their flexibility and capacity of learning many-body interactions \cite{Schutt2017a, Noe2020}.
The vast majority of NN potentials are trained via bottom-up methods \cite{Behler2007, Schutt2017, Gilmer2017, Zhang2018a, Wang2019, Husic2020, Klicpera2020, Klicpera2020b, Qiao2020, Vlachas2021, Jain2021, Ko2021}.
The objective is to match energies and/or forces from a data set, most commonly generated via density functional theory (DFT) for small molecules in vacuum \cite{Ramakrishnan2014}.
Within the data set distribution, state-of-the-art NN potentials have already reached the accuracy limit imposed by DFT, with the test error in predicting potential energy being around 2 orders of
magnitude smaller than the corresponding expected DFT accuracy \cite{Faber2017, Klicpera2020}. 
In the limit of a sufficiently large data set without a distribution shift \cite{Schoenholz2020_adversarial, schwalbekoda2021} with respect to the application domain (potentially generated via active learning approaches \cite{Zhang2019}), remaining deviations of predicted observables from experiments are attributable to uncertainty in DFT simulations \cite{Klicpera2020} --
in line with literature reporting DFT being sensitive to employed functionals \cite{Gillan2016}.
More precise computational quantum mechanics models, e.g. the coupled cluster CCSD(T) method, improve DFT accuracy at the expense of significantly increased computational effort for data set generation \cite{Smith2019, Sauceda2021}.
However, for larger systems such as macromolecules, quantum mechanics computations will remain intractable in the foreseeable future, preventing ab initio data set generation all together.
Thus, the main obstacle in bottom-up learning of NN potentials is currently limited availability of highly precise and sufficiently broad data sets.

Top-down approaches circumvent the need for reliable data-generating simulations.
Leveraging experimental data in the potential optimization process has contributed greatly to the success of classical atomistic \cite{Cornell1995, Oostenbrink2004} and coarse-grained \cite{Marrink2007} (CG) force fields \cite{Frohlking2020}.
Training difficulties have so far impeded a similar approach for NN potentials:
Only recent advances in automatic differentiation (AD) \cite{Baydin2018} software have enabled end-to-end differentiation of molecular dynamics (MD) observables with respect to potential energy parameters  \cite{Schoenholz2020, Doerr2021}, by applying AD through the dynamics of a MD simulation \cite{Ingraham2019, Schoenholz2020, Goodrich2021, Doerr2021}.
This direct reverse-mode AD approach saves all simulator operations on the forward pass to be used during gradient computation on the backward pass, resulting in excessive memory usage. Thus, direct reverse-mode AD for systems with more than hundred particles and a few hundred time steps
is typically intractable \cite{Ingraham2019, Schoenholz2020, Goodrich2021, Doerr2021}. 
Numerical integration of the adjoint equations \cite{Chen2018, Wang2020} represents a memory efficient alternative that requires to save only those atomic configurations that directly contribute to the loss.
However, both approaches backpropagate the gradient through the entire simulation, which dominates computational effort and is prone to exploding gradients, as stated by Ingraham et al. \cite{Ingraham2019} and shown below.

Addressing the call for NN potentials trained on experimental data \cite{Frohlking2020},  we propose the Differentiable Trajectory Reweighting (DiffTRe) method.
DiffTRe offers end-to-end gradient computation and circumvents the need to differentiate through the simulation by combining AD with previous work on MD reweighting schemes \cite{Norgaard2008, Li2011, Carmichael2012, Wang2013a}.
For the common use case of time-independent observables, DiffTRe avoids exploding gradients and reduces computational effort of gradient computations by around 2 orders of magnitude compared to backpropagation through the simulation.
Memory requirements are comparable to the adjoint method.
We showcase the broad applicability of DiffTRe on three numerical test cases: First, we provide insight into the training process on a toy example of ideal gas particles inside a double-well potential. Second, we train the state-of-the-art graph neural network potential DimeNet++ \cite{Klicpera2020, Klicpera2020b} for an atomistic model of diamond from its experimental stiffness tensor. Finally, we learn a DimeNet++ model for CG water based on pressure, as well as radial and angular distribution functions. 
The last example shows how DiffTRe also generalizes bottom-up structural coarse-graining methods such as the iterative Boltzmann inversion \cite{Reith2003} or inverse Monte Carlo \cite{Lyubartsev1995} to many-body correlation functions and arbitrary potentials.
DiffTRe allows to enhance NN potentials with experimental data, which is particularly relevant for systems where bottom-up data is unavailable or not sufficiently accurate.

\section{Results}
\subsection*{Differentiable Trajectory Reweighting}

Top-down potential optimization aims to match the $K$ outputs of a molecular mechanics simulation ${\bm O}$ to experimental observables $\tilde{\bm O}$. 
Therefore, the objective is to minimize a loss function $L(\bm \theta)$, e.g. a mean squared error (MSE)
\begin{equation}
   L(\bm \theta) = \frac{1}{K} \sum_{k=1}^K \left[\langle O_k(U_{\bm \theta}) \rangle - \tilde O_k \right]^2 ,
   \label{eq:definition_mse_loss}
\end{equation}
where $\langle \rangle$ denotes the ensemble average, and $\langle O_k(U_{\bm \theta}) \rangle$ depends on the potential energy $U_{\bm \theta}$ parametrized by $\bm \theta$. We will focus on the case where a MD simulation approximates $\langle O_k(U_{\bm \theta}) \rangle$ - with Monte Carlo \cite{Binder1993} being a usable alternative.
With standard assumptions on ergodicity and thermodynamic equilibrium, the ensemble average $\langle O_k(U_{\bm \theta}) \rangle$ is approximated via a time average
\begin{equation}
    \langle O_k(U_{\bm \theta}) \rangle \simeq \frac{1}{N} \sum_{i=1}^N O_k(\mathbf{S}_i, U_{\bm \theta}) \ ,
\label{eq:time_average}
\end{equation}
where $\{\mathbf{S}_i\}_{i=1}^N$ is the trajectory of the system, i.e. a sequence of $N$ states consisting of particle positions and momenta.
Due to the small time step size necessary to maintain numerical stability in MD simulations, states are highly correlated. Subsampling, i.e. only averaging over every 100th or 1000th state, reduces this correlation in Eq. \eqref{eq:time_average}. 

As the generated trajectory depends on $\bm \theta$, every update of $\bm \theta$ during training would require a re-computation of the entire trajectory. However, by leveraging thermodynamic perturbation theory \cite{Zwanzig1954}, it is possible to re-use decorrelated states obtained via a reference potential $\hat{\bm \theta}$. Specifically, the time average is reweighted to account for the altered state probabilities $p_{\bm \theta}(\mathbf{S}_i)$ from the perturbed potential $\bm \theta$ \cite{Zwanzig1954, Norgaard2008, Li2011}:
\begin{equation}
    \langle O_k(U_{\bm \theta}) \rangle \simeq \sum_{i=1}^N w_i O_k(\mathbf{S}_i, U_{\bm \theta}) \quad \mathrm{with} \quad w_i =
    \frac{p_{\bm \theta}(\mathbf{S}_i) / p_{\hat{\bm \theta}}(\mathbf{S}_i)}{\sum_{j=1}^N p_{\bm \theta}(\mathbf{S}_j) / p_{\hat{\bm \theta}}(\mathbf{S}_j)}.
\label{eq:weights_definition}
\end{equation}
Assuming a canonical ensemble, state probabilities follow the Boltzmann distribution $p_{\bm \theta}(\mathbf{S}_i) \sim e^{-\beta H(\mathbf{S}_i)}$, where $H(\mathbf{S}_i)$ is the Hamiltonian of the state (sum of potential and kinetic energy), $\beta = 1 / (k_B T)$, $k_B$ Boltzmann constant, $T$ temperature. Inserting $p_{\bm \theta}(\mathbf{S}_i)$ into Eq.  \eqref{eq:weights_definition} allows computing weights as a function of $\bm \theta$ (the kinetic energy cancels)
\begin{equation}
    w_i =\frac{e^{-\beta (U_{\bm \theta}(\mathbf{S}_i) - U_{\hat{\bm \theta}}(\mathbf{S}_i)) }}{\sum_{i=j}^N e^{-\beta (U_{\bm \theta}(\mathbf{S}_j) - U_{\hat{\bm \theta}}(\mathbf{S}_j)) }} \ .
    \label{eq:weight_formula}
\end{equation}
For the special case of $\bm \theta = \hat{\bm \theta}$, $w_i = 1 / N$, recovering Eq. \eqref{eq:time_average}.
Note that similar expressions to Eq. \eqref{eq:weight_formula} could be derived for other ensembles, e.g. the isothermal–isobaric ensemble, via respective state probabilities $p_{\bm \theta}(\mathbf{S}_i)$.
In practice, the reweighting ansatz is only applicable given small potential energy differences. For large differences between $\bm \theta$ and $\hat{\bm \theta}$, by contrast, few states dominate the average. In this case, the effective sample size \cite{Carmichael2012}
\begin{equation}
    N_\mathrm{eff} \approx e^{-\sum_{i=1}^N w_i \ln(w_i)}
\end{equation}
is reduced and the statistical error in $\langle O_k(U_{\bm \theta}) \rangle$ increases  (Eq. \eqref{eq:weights_definition}). 

Reweighting can be exploited for two purposes that are linked to speed-ups in the forward and backward pass, respectively: First, reweighting reduces computational effort as decorrelated states from previous trajectories can often be re-used \cite{Carmichael2012}.
Second, and most importantly, reweighting establishes a direct functional relation between $\langle O_k(U_{\bm \theta}) \rangle$ and $\bm \theta$. 
This relation via $\mathbf{w}$ provides an alternative end-to-end differentiable path for computing the gradient of the loss $\nabla_{\bm \theta} L$: Differentiating through the reweighting scheme replaces the backward pass through the simulation.
Leveraging this alternative differentiation path, while managing the effective sample size $N_\mathrm{eff}$, are the central ideas behind the DiffTRe method.

\begin{figure}[t]
    \centering
    \resizebox{\linewidth}{!}{\includegraphics{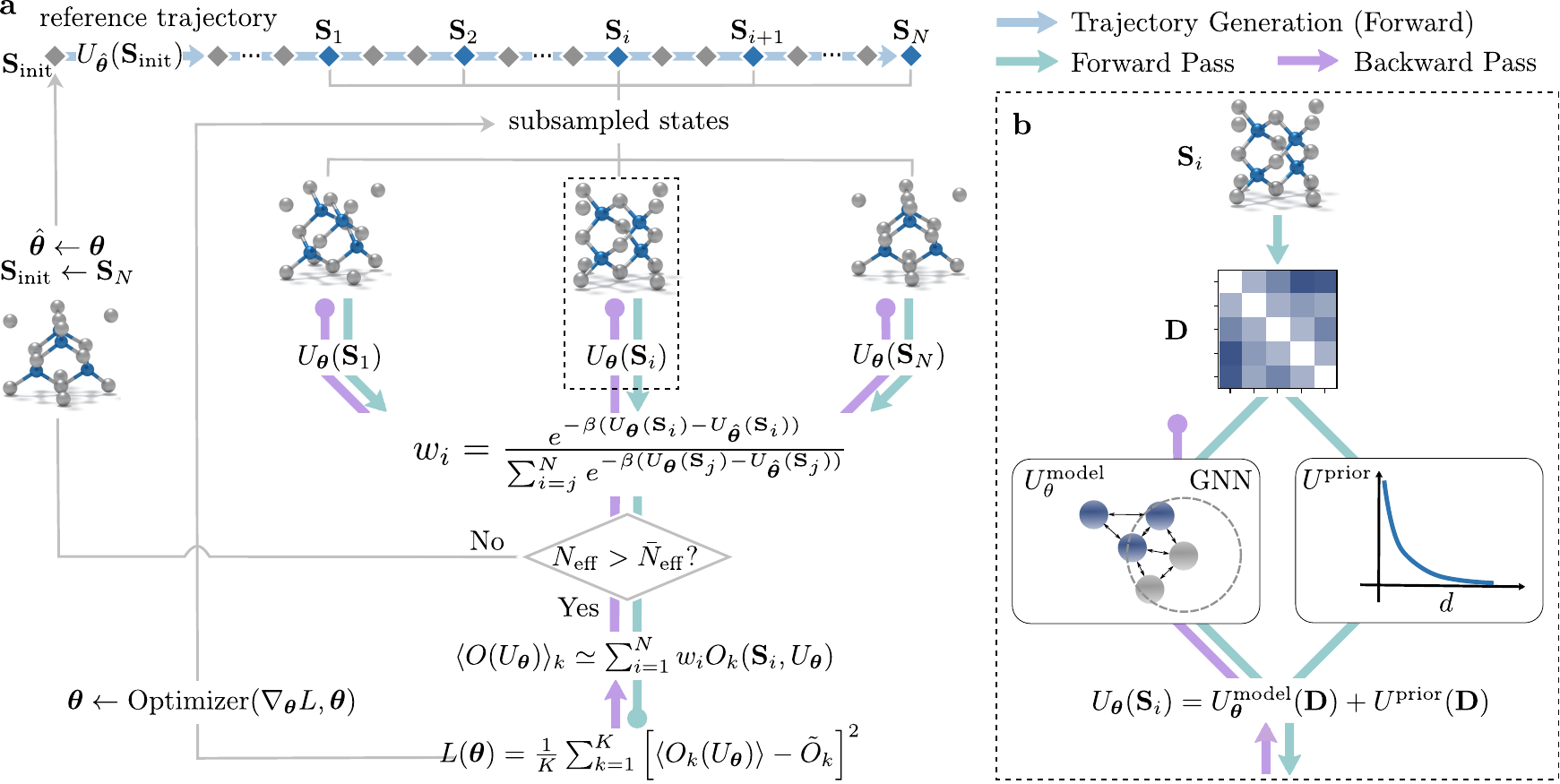}}
    \caption{Differentiable Trajectory Reweighting (DiffTRe). $\mathbf{a}$ Based on an initial state $\mathbf{S}_\mathrm{init}$ and reference potential parameters $\hat{\bm \theta}$, a reference trajectory is generated, of which only subsampled states are retained (blue diamonds), while the majority of visited states are discarded (gray diamonds). For each retained state $\mathbf{S}_i$ (represented by a generic molecular system), the potential energy $U_{\bm \theta}(\mathbf{S}_i)$ and weight $w_i$ are computed under the current potential parameters $\bm \theta$. $w_i$ allow computation of reweighted observables $\langle O_k(U_{\bm \theta}) \rangle$, the loss $L(\bm \theta)$, its gradient $\nabla_{\bm \theta} L$ and subsequently, updating $\bm \theta$ via the optimizer. The updating procedure is repeated until the effective sample size $N_\mathrm{eff} < \bar N_\mathrm{eff}$, at which point a new reference trajectory needs to be generated starting from the last sampled state $\mathbf{S}_N$.
    $\mathbf{b}$ Computation of $U_{\bm \theta}(\mathbf{S}_i)$ from the pairwise distance matrix $\mathbf{D}$, which is fed into the learnable potential $U_{\bm \theta}^\mathrm{model}$ (e.g. a graph neural network - GNN) and $U^\mathrm{prior}$ (e.g. a pairwise repulsive potential). 
    }
    \label{fig:DiffTRe_scheme}
\end{figure}

The workflow of the DiffTRe algorithm consists of the following steps:
First, an initial reference trajectory is generated from the canonical ensemble, e.g. via a stochastic or deterministic thermostat, from an initial state $\mathbf{S}_\mathrm{init}$ and reference potential $\hat{\bm \theta}$ (Fig. \ref{fig:DiffTRe_scheme} $\mathbf{a}$). Initial equilibration states are disregarded and the following states are subsampled yielding decorrelated states $\{\mathbf{S}_i\}_{i=1}^N$. Together with their reference potential energies $\{U_{\hat{\bm \theta}}(\mathbf{S}_i)\}_{i=1}^N$, these states are saved for re-use during reweighting. 
In the next step, the reweighting scheme is employed to compute $\nabla_{\bm \theta} L$ with respect to current parameters $\bm \theta$, where initially $\bm \theta = \hat{\bm \theta}$. An optimizer subsequently uses $\nabla_{\bm \theta} L$ to improve $\bm \theta$. 
This procedure of reweighting, gradient computation and updating is repeated as long as the statistical error from reweighting is acceptably small, i.e. $N_\mathrm{eff}$ is larger than a predefined $\bar N_\mathrm{eff}$. As soon as $N_\mathrm{eff} < \bar N_\mathrm{eff}$, a new reference trajectory needs to be sampled using the current $\bm \theta$ as the new $\hat{\bm \theta}$.
At least one ${\bm \theta}$ update per reference trajectory is ensured because initially $N_\mathrm{eff} = N$. Using the last generated state $\mathbf{S}_N$ as $\mathbf{S}_\mathrm{init}$ for the next trajectory counteracts overfitting to a specific initial configuration. Additionally, $p_{\hat{\bm \theta}}(\mathbf{S}_\mathrm{init}$) is reasonably high when assuming small update steps, reducing necessary equilibration time for trajectory generation. Saving only $\{\mathbf{S}_i\}_{i=1}^N$ and $\{U_{\hat{\bm \theta}}(\mathbf{S}_i)\}_{i=1}^N$ from the simulation entails low memory requirements similar to the adjoint method. DiffTRe assumes that deviations in predicted observables are attributable to an inaccurate potential $U_{\bm \theta}$ rather than statistical sampling error. Accordingly, $N$ and the subsampling ratio $n$ need to be chosen to yield sufficiently small statistical error. Optimal values for $N$ and $n$ depend on the specific system, target observables and the thermodynamic state point.

Computation of $\nabla_{\bm \theta} L$ via reverse-mode AD through the reweighting scheme comprises a forward pass starting with computation of the potential $U_{\bm \theta}(\mathbf{S}_i)$ and weight $w_i$ for each $\mathbf{S}_i$ (Eq. \eqref{eq:weight_formula} ; Fig. \ref{fig:DiffTRe_scheme} $\mathbf{a}$). Afterwards, reweighted observables $\langle O_k(U_{\bm \theta}) \rangle$ (Eq. \eqref{eq:weights_definition}) and the resulting loss $L(\bm \theta)$ (Eq. \eqref{eq:definition_mse_loss}) are calculated.
The corresponding backward pass starts at $L(\bm \theta)$ and stops at parameters $\bm \theta$ in the potential energy computation $U_{\bm \theta}(\mathbf{S}_i)$. The differentiation path defined by the reweighting ansatz is therefore independent of the trajectory generation.

Evaluation of $U_{\bm \theta}(\mathbf{S}_i)$ (Fig. \ref{fig:DiffTRe_scheme} $\mathbf{b}$) involves computing the pairwise distance matrix $\mathbf{D}$ from atom positions of $\mathbf{S}_i$, that are fed into a learnable potential $U_{\bm \theta}^\mathrm{model}$ and a prior potential $U^\mathrm{prior}$. Both potential components are combined by adding the predicted potential energies
\begin{equation}
    U_{\bm \theta}(\mathbf{S}_i) = U^\mathrm{model}_{\bm \theta}(\mathbf{D}) + U^\mathrm{prior}(\mathbf{D}) .
\end{equation}
In subsequent examples of diamond and CG water, $U_{\bm \theta}^\mathrm{model}$ is a graph neural network operating iteratively on the atomic graph defined by $\mathbf{D}$. $U^\mathrm{prior}$ is a constant potential approximating a-priori known properties of the system, such as the Pauli exclusion principle (e.g. Eq. \eqref{eq:prior_repulsion}).
Augmenting NN potentials with a prior is common in the bottom-up coarse-graining literature \cite{Zhang2018a, Husic2020} to provide qualitatively correct behavior in regions of the potential energy surface (PES) not contained in the data set, but reachable by the CG model. By contrast, DiffTRe does not rely on pre-computed data sets. Rather, the prior serves to control the data (trajectory) generation in the beginning of the optimization.
Additionally, $U^\mathrm{prior}$ reformulates the problem from learning $U_{\bm \theta}^\mathrm{model}$ directly to learning the difference between $U^\mathrm{prior}$ and the optimal potential given the data \cite{Husic2020}. 
A well-chosen $U^\mathrm{prior}$ therefore represents a physics-informed initialization accelerating training convergence.
Suitable $U^\mathrm{prior}$ can often be found in the literature: Classical force fields such as AMBER \cite{Cornell1995} and MARTINI \cite{Marrink2007} define reasonable interactions for bio-molecules and variants of the Embedded Atom Model \cite{Daw1984} (EAM) provide potentials for metals and alloys.
Note that $U^\mathrm{prior}$ is not a prior in the Bayesian sense providing a pervasive bias on learnable parameters in the small data regime. If $U^\mathrm{prior}$ is in contradiction with the data, $U_{\bm \theta}^\mathrm{model}$ will correct for $U^\mathrm{prior}$ as a result of the optimization. 
In the next section, we further illustrate for a toy problem the interplay between prior, gradients and the learning process in DiffTRe, and provide a comparison to direct reverse-mode AD through the simulation.

\subsection*{Double-well toy example}

\begin{figure}[t!]
    \centering
    \resizebox{\linewidth}{!}{\includegraphics{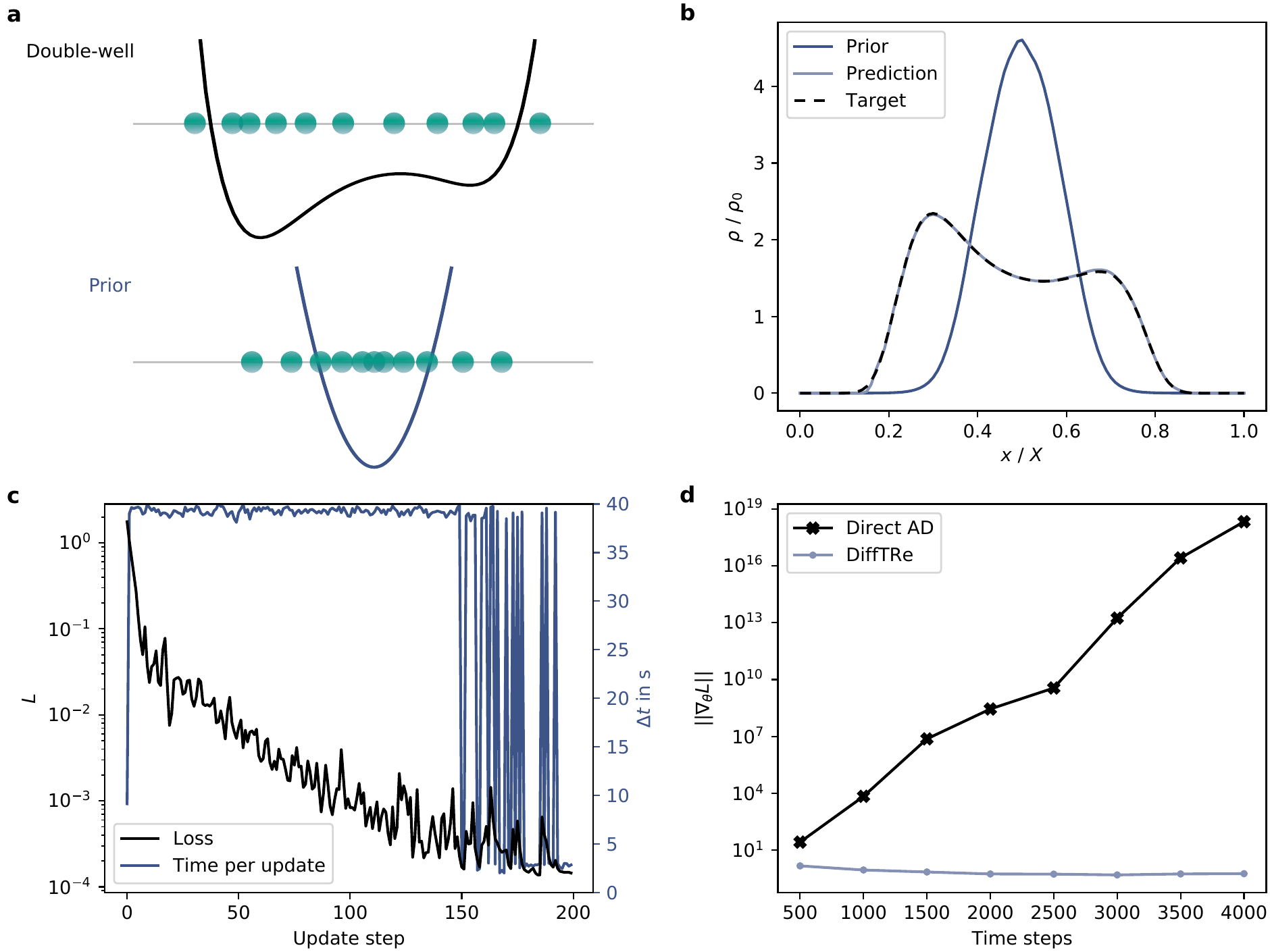}}
    \caption{Double-well toy example. $\mathbf{a}$ Sketch of the double-well and prior potential with corresponding example states of ideal gas particles (green circles).
    The learned potential results in a normalized density $\rho / \rho_0$ (over the normalized position $x / X$) that matches the target closely ($\mathbf{b}$).
    Successful learning is reflected in the loss curve $L$, where significant reduction in wall-clock time per parameter update $\Delta t$ towards the end of the optimization is achieved through re-using previously generated trajectories ($\mathbf{c}$).
    Gradients computed via DiffTRe have constant magnitudes while gradients obtained from direct reverse-mode automatic differentiation through the simulation suffer from exploding gradients for longer trajectories ($\mathbf{d}$).}
    \label{fig:results_double_well}
\end{figure}

We consider ideal gas particles at a temperature $k_B T = 1$ trapped inside a one-dimensional double-well potential (Fig. \ref{fig:results_double_well} $\mathbf{a}$) parametrized by
\begin{equation}
    U(x) = k_B T * \left[2500 (x - 0.5)^6 - 10 (x - 0.55)^2\right] \ .
\end{equation}
The goal is to learn $\bm \theta$ such that $U_{\bm \theta}(x) = U_{\bm \theta}^\mathrm{model}(x) + U^\mathrm{prior}(x)$ matches $U(x)$. We select a cubic spline as $U_{\bm \theta}^\mathrm{model}$, which acts as a flexible approximator for twice continuously differentiable functions. The cubic spline is parametrized via the potential energy values of 50 control points $\{x_j, U_j\}_{j=1}^{50}$ evenly distributed over $x \in [0,1]$. Analogous to NN potentials in subsequent problems, we randomly initialize $U_j \sim \mathcal{N}(0,0.01^2 k_B T)$. Initializing $U_j = 0$ leads to largely identical results in this toy problem. The harmonic single-well potential $U^\mathrm{prior}(x) = \lambda (x - 0.5)^2$, with scale $\lambda = 75$, encodes the prior knowledge that particles cannot escape the double-well. We choose the normalized density profile $\rho(x) / \rho_0$ of ideal gas particles as the target observable. The resulting loss function is
\begin{equation}
    L = \frac{1}{K} \sum_{k=1}^K \left(\frac{\langle \rho(x_k) \rangle}{\rho_0} - \frac{\tilde \rho(x_k)}{\rho_0} \right)^2 \ ,
    \label{eq:double_well_loss}
\end{equation}
where $\rho(x)$ is discretized via $K$ bins. $\langle \rho(x_k) \rangle$ are approximated based on $N=10000$ states after skipping 1000 states for equilibration, where a state is retained every 100 time steps.
We minimize Eq. \eqref{eq:double_well_loss} via an Adam \cite{Kingma2015} optimizer with learning rate decay. For additional DiffTRe and simulation parameters, see Supplementary Method 1.1.

Initially, $\rho / \rho_0$ resulting from $U^\mathrm{prior}(x)$ deviates strongly from the target double-well density (Fig. 
\ref{fig:results_double_well} $\mathbf{b}$). 
The loss curve illustrates successful optimization over 200 update steps (Fig. \ref{fig:results_double_well} $\mathbf{c}$).
The wall-clock time per parameter update $\Delta t$ clearly shows two distinct levels: At the start of the optimization, update steps are rather large, significantly reducing $N_\mathrm{eff}$. Hence a new reference trajectory generation is triggered with each update (average $\Delta t \approx 39.2$ s). Over the course of the simulation, updates of $U_{\bm \theta}^\mathrm{model}(x)$ become smaller and reference trajectories are occasionally re-used (average $\Delta t \approx 2.76$ s).
After optimization, the  target density is matched well. The learned potential energy function $U_{\bm \theta}(x)$ recovers the data-generating potential $U(x)$ (Supplementary Fig. 1 $\mathbf{a}$); thus, other thermodynamic and kinetic observables will match reference values closely. However, this conclusion does not apply in realistic applications, where learned potentials are in general not unique \cite{Noid2013} due to the limited number of target observables that can be considered in practice.

The effect of $U^\mathrm{prior}$ on the training process is twofold: 
First, by encoding prior knowledge, it simplifies convergence, as $U_{\bm \theta}^\mathrm{model}(x)$ only needs to adapt the single-well prior instead of learning large energy barriers from scratch.
Second, $U^\mathrm{prior}$ also impacts the information content of the gradient by controlling the generation of trajectories in the beginning of the optimization (Eq. \eqref{eq:DiffTRe_gradient}). The local support of the cubic spline allows analyzing this relation empirically (Supplementary Fig. 2): The gradient is non-zero only in regions of the PES that are included in the reference trajectory. Hence, other regions of the PES are not optimized despite delivering a non-zero contribution to the loss.
A well-chosen prior potential should therefore yield trajectories that are as close as possible to trajectories sampled from the true potential.
However, satisfactory learning results can be obtained for a sensible range of prior scales (Supplementary Fig. 3). 

We study the robustness of our results by varying the random seed that controls the initialization of the spline as well as the initial particle positions and velocities. 
Results from the variation study in Supplementary Fig. 4 demonstrate that the predicted $\rho(x) / \rho_0$ is robust to random initialization. The corresponding $U_{\bm \theta}(x)$ exhibits some variance at the left well boundary, mirroring difficult training in this region due to vanishing gradients for vanishing predicted densities (Supplementary Fig. 2) and minor influence of the exact wall position on the resulting density profile (Supplementary Fig. 4 $\mathbf{a}$).

For comparison, we have implemented gradient computation via direct reverse-mode AD through the simulation. This approach clearly suffers from the exploding gradients problem (Fig. \ref{fig:results_double_well} $\mathbf{d}$): The gradient magnitude increases exponentially as a function of the simulation length. Without additional modifications (e.g. as implemented by Ingraham et al. \cite{Ingraham2019}), these gradients are impractical for longer trajectories. By contrast, gradients computed via DiffTRe show constant magnitudes irrespective of the simulation length.

To measure the speed-up over direct reverse-mode AD empirically, we simulate the realistic case of an expensive potential by substituting the numerically inexpensive spline by a fully-connected neural network with 2 hidden layers and 100 neurons each. We measure speed-ups of $s_g = 486$ for gradient computations and $s = 3.7$ as overall speed-up per update when a new reference trajectory is sampled. However, these values are rather sensitive to the exact computational and simulation setup.
Memory overflow in the direct AD method constrained trajectory lengths to 10 retained states and a single state for equilibration (a total of 1100 time steps). Measuring speed-up for one of the real-world problems below would be desirable, but is prevented by the memory requirements of direct AD. 

The measured speed-up values are in line with theoretical considerations: While direct AD backpropagates through the whole trajectory generation, DiffTRe only differentiates through the potential energy computation of decorrelated states $\{\mathbf{S}_i\}_{i=1}^N$ (Fig. \ref{fig:DiffTRe_scheme}). 
From this algorithmic difference, we expect speed-up values that depend on the subsampling ratio $n$, the number of skipped states during equilibration $N_\mathrm{equilib}$ and the cost multiple of backward passes with respect to forward passes $G$ (details in Supplementary Method 2)
\begin{equation}
    s_g \sim G n \left(1 + N_\mathrm{equilib} / N \right) \ ; \quad s \sim G + 1 \ .
    \label{eq:speedup_estimate}
\end{equation}
For this toy example setup, the rule-of-thumb estimates in Eq. \eqref{eq:speedup_estimate} yield $s_g = 330$ and $s=4$, agreeing with the measured values.
In the next sections, we showcase the effectiveness of DiffTRe in real-world, top-down learning of NN potentials.

\subsection*{Atomistic model of Diamond}
\begin{figure}[t!]
    \centering
    \resizebox{\linewidth}{!}{\includegraphics{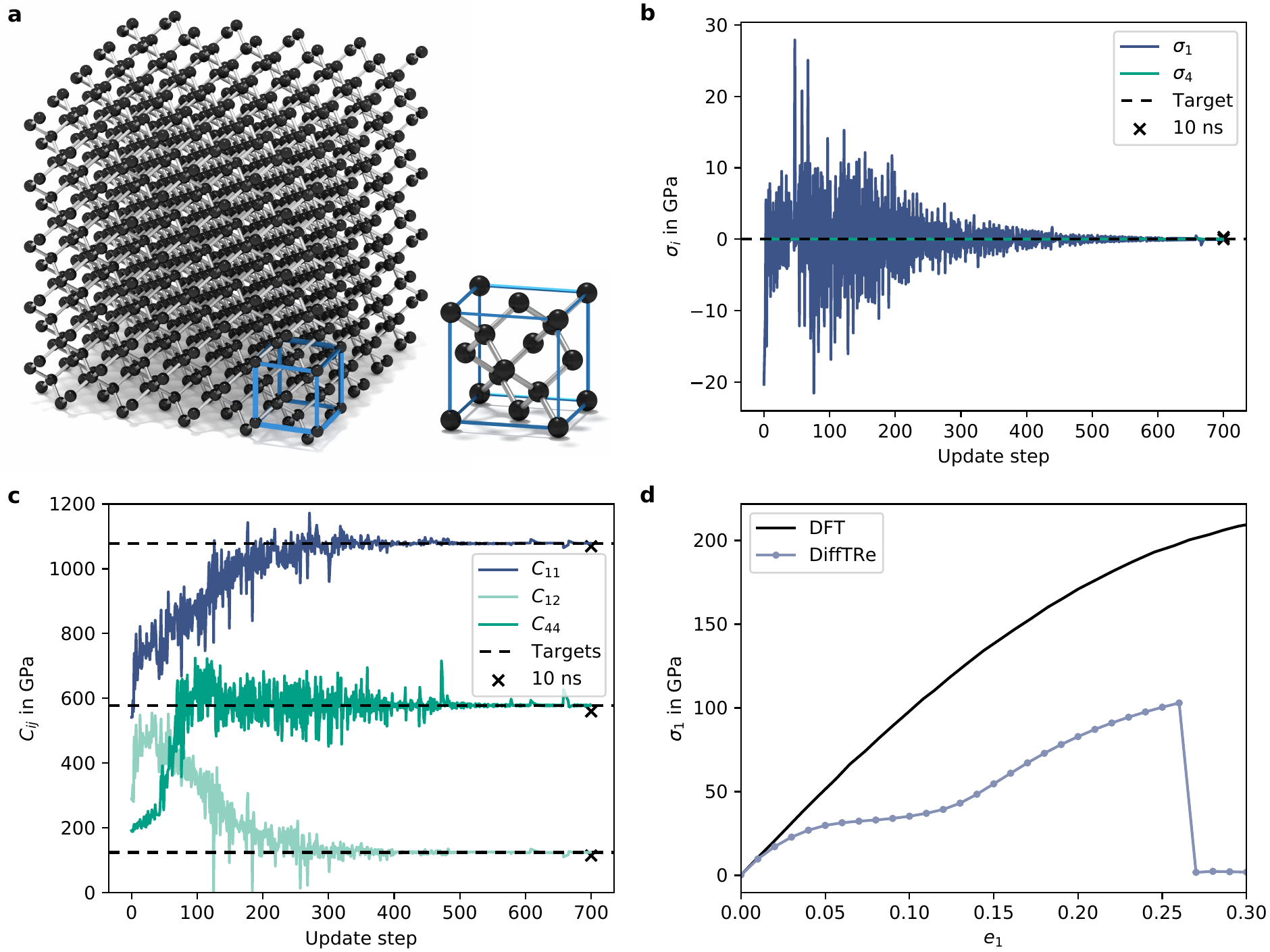}}
    \caption{Atomistic model of diamond. The simulation box consists of 5 diamond unit cells in each direction, whose primary crystallographic directions $[1,0,0]$, $[0,1,0]$ and $[0,0,1]$ are aligned with the x, y and z axes of the simulation box ($\mathbf{a}$). Stress $\sigma_{i}$ ($\mathbf{b}$) and stiffness values $C_{ij}$ ($\mathbf{c}$) converge to their respective targets during the optimization. These results are robust to long simulation runs of 10 ns (marked with crosses). 
    The stress-strain curve over normal natural strains $e_1$ agrees with density functional theory (DFT) data \cite{Jensen2015} for medium-sized strains ($e_1 <= 0.02$), but deviates for large strains due to limited extrapolation capabilities of neural network potentials ($\mathbf{d}$).}
    \label{fig:results_diamond}
\end{figure}

To demonstrate applicability of DiffTRe to solids on the atomistic scale, we learn a DimeNet++ \cite{Klicpera2020b} potential for diamond from its experimental elastic stiffness tensor $\mathbf{C}$. Due to symmetries in the diamond cubic crystal, $\mathbf{C}$ only consists of three distinct stiffness moduli $\tilde C_{11} = 1079$ GPa, $\tilde C_{12} = 124$ GPa and $\tilde C_{44} = 578$ GPa \cite{McSkimin1972a} (in Voigt notation). Additionally, we assume the crystal to be in a stress-free state $\bm \sigma = \mathbf{0}$ for vanishing infinitesimal strain $\bm \epsilon = \mathbf{0}$. These experimental data define the loss
\begin{equation}
    L = \frac{\gamma_{\bm \sigma}}{9} \sum_{i=1, j=1}^{i=3, j=3}\sigma_{ij}^2 + \frac{\gamma_{\mathbf{C}}}{3} \left( (C_{11} - \tilde C_{11})^2 + (C_{12} - \tilde C_{12})^2 + (C_{44} - \tilde C_{44})^2 \right) \ ,
\end{equation}
where loss weights $\gamma_{\bm \sigma}$ and $\gamma_{\mathbf{C}}$ counteract the effect of different orders of magnitude of observables.
To demonstrate learning, we select the original Stillinger-Weber potential \cite{Stillinger1985} parametrized for silicon as $U^\mathrm{prior}$. We have adjusted the length and energy scales to $\sigma_\mathrm{SW} = 0.14 \ \mathrm{nm}$ and $\epsilon_\mathrm{SW} = 200 \ \mathrm{kJ / mol}$, reflecting the smaller size of carbon atoms. 
We found learning to be somewhat sensitive to $U^\mathrm{prior}$ in this example because weak prior choices can lead to unstable MD simulations.
Simulations are run with a cubic box of size $L \approx 1.784$ nm containing 1000 carbon atoms (Fig. \ref{fig:results_diamond} $\mathbf{a}$) to match the experimental density ($\rho=3512 \ \mathrm{kg / m^3}$)\cite{McSkimin1972a} exactly. The temperature in the experiment ($T = 298.15$ K \cite{McSkimin1972a}) determines the simulation temperature. Each trajectory generation starts with 10 ps of equilibration followed by 60 ps of production, where a decorrelated state is saved every 25 fs. We found these trajectories to yield observables with acceptably small statistical noise. The stress tensor $\bm \sigma$ is computed via Eq. \eqref{eq:stress_tensor} and the stiffness tensor $\mathbf{C}$ via the stress fluctuation method (Eq. \eqref{eq:stress_fluctuation_method}). Further details are summarized in Supplementary Method 4. 

Fig. \ref{fig:results_diamond} visualizes convergence of the stress ($\mathbf{b}$) and stiffness components ($\mathbf{c}$).
Given that the model is only trained on rather short trajectories, we test the trained model on a trajectory of 10 ns length to ensure that the model neither overfitted to initial conditions nor drifts away from the targets. 
The resulting stress and stiffness values $\sigma_1 = 0.29$ GPa, $\sigma_4 = 0.005$ GPa, $C_{11} = 1070$ GPa,  $C_{12} = 114$ GPa, and  $C_{44} = 560$ GPa are in good agreement with respective targets. These results could be improved by increasing the trajectory length, which reduces statistical sampling errors.
The corresponding inverse stress-strain relation is given by the compliance tensor $\mathbf{S} = \mathbf{C}^{-1}$, which can be constructed from by Young's modulus $E = 1047$ GPa, shear modulus $G = 560$ GPa and Poisson's ratio $\nu = 0.097$.
The training loss curve and wall-clock time per update $\Delta t$ are displayed in Supplementary Fig. 5 $\mathbf{a}$. 

Computing the stress-strain curve (Supplementary Fig. 5 $\mathbf{b}$) from the trained model in the linear regime ($\epsilon_i < 0.005$) verifies that computing $\mathbf{C}$ via Eq. \eqref{eq:stress_fluctuation_method} yields the same result as explicitly straining the box and measuring stresses. Additionally, this demonstrates that the DimeNet++ potential generalizes from the training box ($\bm \epsilon = 0$) to boxes under small strain. 
We also strained the box beyond the linear regime, creating a distribution shift \cite{Schoenholz2020_adversarial, schwalbekoda2021}, to test generalization to unobserved state points.
The predicted stress-strain curve in Fig. \ref{fig:results_diamond} $\mathbf{d}$ shows good agreement with DFT data \cite{Jensen2015} for medium-sized natural strains $e_1 = \log(1 + \epsilon_1) < 0.02$. For large strains the deviation quickly increases,  including an early fracture. These incorrect predictions of the learned potential are due to limited extrapolation capacities of NN potentials: States under large strain are never encountered during training, leading to large uncertainty in predicted forces. 
Incorporating additional observables linked to states of large strain into the optimization, such as the point of maximum stress, should improve predictions.

To test the trained DimeNet++ potential on held-out observables, we compute the phonon density of states (PDOS). The predicted PDOS deviates from the experiment \cite{Dolling1966}, analogous to a Stillinger-Weber potential optimized for diamond \cite{Barnard2002} (Supplementary Fig. 5 $\mathbf{c}$). The evolution of the predicted PDOS over the course of the optimization is shown in Supplementary Fig. 5 $\mathbf{d}$. Deviations of held-out observables are expected given that top-down approaches allow learning potentials that are consistent with target experimental observables, but lack theoretical convergence guarantees of bottom-up schemes (in the limit of a sufficiently large data set and a sufficiently expressive model) \cite{Noid2013}. In principle, we expect sufficiently expressive top-down models to converge to the true potential in the limit of an infinite number of matched target observables. In practice however, many different potentials can reproduce a sparse set of considered target observables, rendering the learned potential non-unique \cite{Noid2013}.
In this particular example, we show that many different potentials can reproduce the target stress and stiffness, but predict different PDOSs:
While predicted stress and stiffness values are robust to random initialization of NN weights and initial particle velocities within the statistical sampling error, the corresponding predicted PDOSs vary to a great extent (Supplementary Fig. 6).
Incorporating additional observables more closely connected to phonon properties into the loss function could improve the predicted PDOS.

\subsection*{Coarse-grained water model}
\begin{figure}[t!]
    \centering
    \resizebox{\linewidth}{!}{\includegraphics{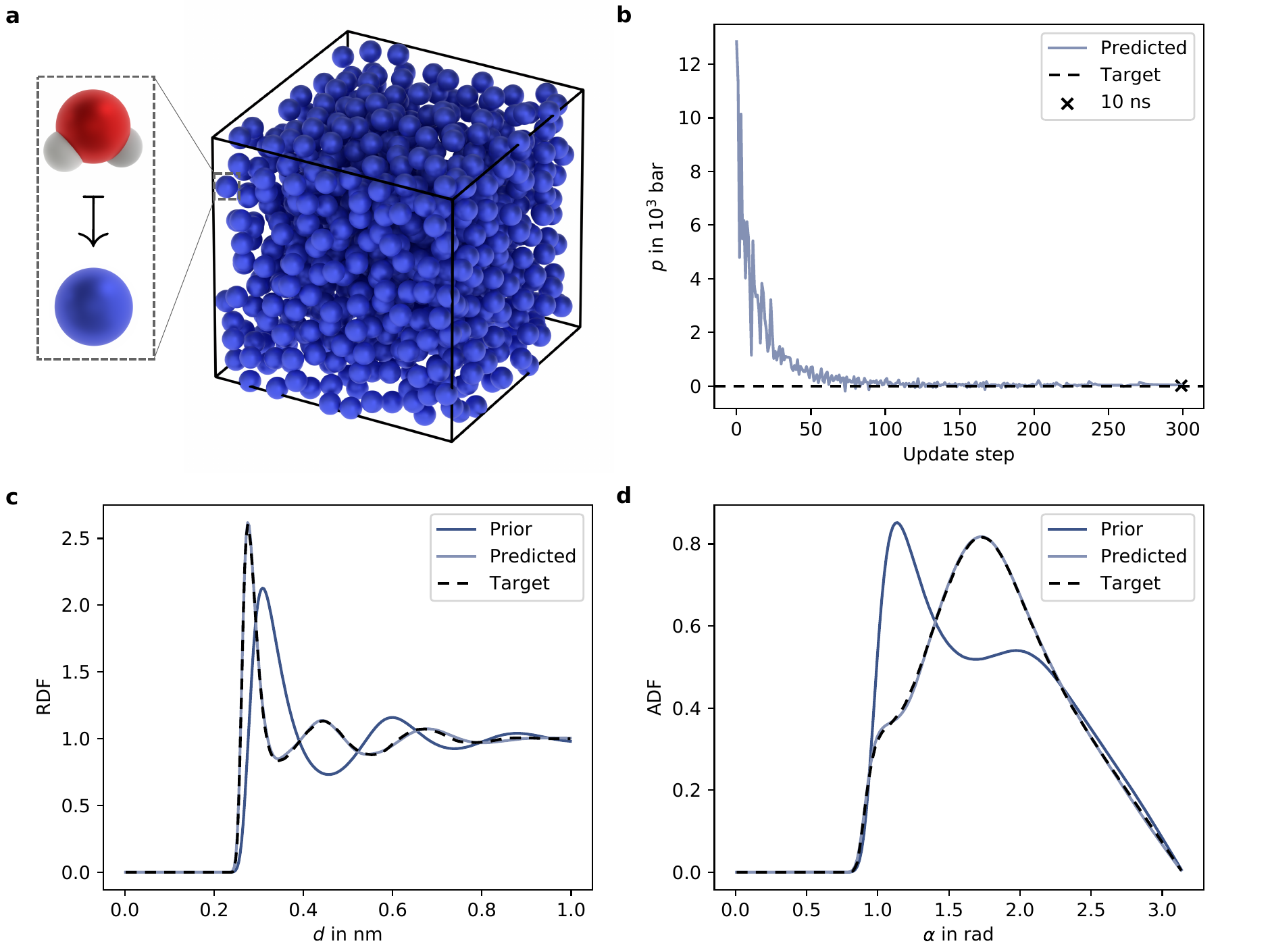}}
    \caption{Coarse-grained model of water. Coarse-grained particles representing water molecules are visualized as blue balls in the simulation box ($\mathbf{a}$). The pressure $p$ converges quickly towards its target of 1 bar during optimization and the subsequent 10 ns simulation (black cross; $p \approx 12.9$ bar) verifies the result ($\mathbf{b}$). Over a 10 ns simulation, the learned potential reconstructs the experimental radial distribution function (RDF) and angular distribution function (ADF) well ($\mathbf{c} - \mathbf{d}$).}
    \label{fig:results_water}
\end{figure}

% substitute RDF by g(r) and ADF by ?? Use g(d) for consistency with figure 1
Finally, we learn a DimeNet++ potential for CG water. Water is a common benchmark problem due to its relevance in bio-physics simulations and its pronounced 3-body interactions, which are challenging for classical potentials \cite{Scherer2018}. We select a CG-mapping, where each CG particle is centered at the oxygen atom of the corresponding atomistic water molecule (Fig. \ref{fig:results_water} $\mathbf{a}$). This allows using experimental oxygen-oxygen radial (RDF) and angular distribution functions (ADF) as target observables. Given that the reference experiment \cite{Soper2008} was carried out at ambient conditions ($T = 296.15$ K), we can additionally target a pressure $\tilde p = 1$ bar. Hence, we minimize
\begin{equation}
    L = \frac{1}{G} \sum_{g=1}^G (RDF(d_g) - \tilde{RDF}(d_g))^2 + \frac{1}{M} \sum_{m=1}^M (ADF(\alpha_m) - \tilde{ADF}(\alpha_m))^2 + \gamma_p(p - \tilde p)^2 \ .
\end{equation}
As the prior potential, we select the repulsive term of the Lennard-Jones potential 
\begin{equation}
    U^\mathrm{prior}(d) = \epsilon_R \left(\frac{\sigma_R}{d}\right)^{12} \ .
    \label{eq:prior_repulsion}
\end{equation}
Drawing inspiration from atomistic water models, we have chosen the length scale of the SPC \cite{Berendsen1981} water model as $\sigma_R = 0.3165$ nm as well as a reduced energy scale of $\epsilon_R = 1$ kJ / mol to counteract the missing Lennard-Jones attraction term in Eq. \eqref{eq:prior_repulsion}. We build a cubic box of length 3 nm with 901 CG particles, implying a density of $\rho=998.28$ g/l, to match the experimental water density of $\rho=997.87$ g/l at 1 bar. Trajectory generation consists of 10 ps of equilibration and 60 ps of subsequent production, where a decorrelated state is saved every 0.1 ps.
For additional details, see Supplementary Method 2.3.

Fig. \ref{fig:results_water} $\mathbf{b}$ - $\mathbf{d}$ displays properties predicted by the final trained model during a 10 ns production run:
DiffTRe is able to train a DimeNet++ potential that simultaneously matches experimental oxygen RDF, ADF and pressure to the line thickness.
The evolution of predicted RDFs and ADFs as well as the loss and wall-clock times per update are displayed in Supplementary Fig. 7 $\mathbf{a}$ - $\mathbf{c}$. The learning process is robust to weak choices of $U^\mathrm{prior}$: DiffTRe is able to converge to the same prediction quality as with the reference prior even if $\sigma_R$ is misestimated by $\pm 0.1$ nm (approximately $\pm 30\%$) compared to the classical SPC water model (Supplementary Fig. 8 $\mathbf{a} - \mathbf{b}$). This represents a large variation given that within common atomistic water models, $\sigma_R$ varies by less than $0.5\%$ \cite{Wu2006}.

To test the learned potential on held-out observables, we compute the tetrahedral order parameter $q$ \cite{Errington2001}  and the self-diffusion coefficient $D$.
$q \approx 0.569$ matches the experimental value of $\tilde q = 0.576$ closely. This is expected as $q$ considers the structure of 4 nearest neighbor particles, which is closely related to the ADF. The learned CG water model predicts a larger self-diffusion coefficient than were experimentally measured ($D = 10.91 \ \mu \mathrm{m}^2/\mathrm{ms}$ vs. $\tilde D = 2.2 \ \mu \mathrm{m}^2/\mathrm{ms}$) \cite{Mills1973}.
With the same simulation setup, a single-site tabulated potential parametrized via iterative Boltzmann inversion \cite{Reith2003} with pressure correction \cite{Reith2003, Wang2009} predicts $D = 14.15 \ \mu \mathrm{m}^2/\mathrm{ms}$. These results are in line with the literature: Due to smoother PESs, CG models exhibit accelerated dynamical processes compared to atomistic models \cite{Noid2013}. For CG water models specifically, diffusion coefficients decrease with increasing number of interaction sites \cite{Matysiak2008}. In this context, the decreasing diffusion coefficients over the course of the optimization (Supplementary Fig. 7 $\mathbf{d}$) could indicate that $U_{\bm \theta}$ acts effectively as a single-site model in the beginning, while learning 3-body interactions during the optimization casts $U_{\bm \theta}$ more similar to multi-site CG models.
Obtained results are robust to random initialization of NN weights and initial particle velocities, both for predicted target (Supplementary Fig. 8 $\mathbf{c} - \mathbf{d}$) and held-out observables ($D= 10.93 \pm 0.20 \ \mu \mathrm{m}^2/\mathrm{ms}$).

The accuracy of predicted 2 and 3-body interactions (Fig. \ref{fig:results_water} $\mathbf{c}$ and $\mathbf{d}$) showcases the potency of graph neural network potentials in top-down molecular modeling: 
Capturing 3-body interactions is essential for modeling water given that pair potentials trained via force matching fail to reproduce both RDF and ADF of the underlying high-fidelity model \cite{Scherer2018}.
Other top-down CG water models with simple functional form tend to deviate from the experimental RDF \cite{Molinero2009, Chan2019c}.
Deviations from experimental structural properties, albeit smaller in size, also arise in DFT simulations \cite{Distasio2014, Gillan2016}, limiting the accuracy of bottom-up trained NN potentials \cite{Zhang2018a}.

\section{Discussion}
In this work, we demonstrate numerically efficient learning of NN potentials from experimental data. Main advantages of our proposed DiffTRe method are its flexibility and simplicity: DiffTRe is applicable to solid and fluid materials, coarse-grained and atomistic models, thermodynamic, structural and mechanical properties, as well as potentials of arbitrary functional form. To apply DiffTRe, practitioners only need to set up a MD simulation with corresponding observables and a loss function, while gradients are computed conveniently in an end-to-end fashion via AD. The demonstrated speed-ups and limited memory requirements promote application to larger systems.

Without further adaptations, DiffTRe can also be applied as a bottom-up model parametrization scheme. In this case, a high-fidelity simulation, rather than an experiment, provides target observables. For CG models, DiffTRe generalizes structural coarse-graining schemes such as iterative Boltzmann inversion \cite{Reith2003} or Inverse Monte Carlo \cite{Lyubartsev1995}. DiffTRe overcomes the main limitations of these approaches: First, structural coarse-graining is no longer restricted to one-dimensional potentials and matching many-body correlation functions (e.g. ADFs) is therefore feasible. Second, the user can integrate additional observables into the optimization without relying on hand-crafted iterative update rules, for instance for pressure-matching \cite{Reith2003, Wang2009}.
This is particularly useful if an observable needs to be matched precisely (e.g. pressure in certain multiscale simulations \cite{Thaler2020}).
Matching many-body correlation functions will likely allow structural bottom-up coarse-graining to take on significance within the new paradigm of many-body CG potentials \cite{Zhang2018a, Wang2019, Husic2020}. 

For practical application of DiffTRe, a few limitations need to be considered.
The reweighting scheme renders DiffTRe invariant to the sequence of states in the trajectory. Hence, dynamical properties cannot be employed as target observables. 
Additionally, the NN potential test cases considered in this work required a reasonably chosen prior potential.
Lastly, two distinct sources of overfitting when learning from experimental data for a single system need to be accounted for \cite{Frohlking2020}: 
To avoid overfitting to a specific initial state, DiffTRe uses a different initial state for each reference trajectory. 
Moreover, increasing the system size and trajectory length ensures representative reference trajectories.
Irrespective of overfitting, generalization to different systems, observables and thermodynamic state points remains to be addressed, for instance via training on multi-systemic experimental data sets.
To this end, an in-depth assessment of out-of-sample properties of top-down learned NN potentials is required.

From a machine learning (ML) perspective, DiffTRe belongs to the class of end-to-end differentiable physics approaches \cite{Belbute-Peres2018, Innes2019, Hu2019}. These approaches are similar to reinforcement learning in that the target outcome of a process (here a MD simulation) represents the data. A key difference is availability of gradients through the process, allowing for efficient training.
Differentiable physics approaches, increasingly popular in control applications \cite{Degrave2019, Wang2020, Holl2020, Schafer2020}, enable direct training of the ML model via the physics simulator, advancing the ongoing synthesis of ML and physics-based methods.

Finally, the combination of bottom-up and top-down approaches for learning NN potentials, i.e. considering information from both the quantum and macroscopic scale, represents an exciting avenue for future research. For top-down approaches, pre-training NN potentials on bottom-up data sets can serve as a sensible extrapolation for the PES in areas unconstrained by the experimental data.
In DiffTRe, a pre-trained model could also circumvent the need for a prior potential.
Bottom-up trained NN potentials, on the other hand, can be enriched with experimental data, which enables targeted refinement of the potential. 
This is particularly helpful for systems in which DFT accuracy is insufficient or generation of a quantum mechanical data set is computationally intractable.

\section{Methods}

\subsection*{Differentiable histogram binning}
To obtain an informative gradient $\frac{\partial L}{\partial \bm \theta}$, predicted observables need to be continuously differentiable.
However, many common observables in MD, including density and structural correlation functions, are computed by discrete histogram binning. 
To obtain a differentiable observable, the (discrete) Dirac function used in binning can be approximated by a narrow Gaussian probability density function (PDF) \cite{Wang2020}. Similarly, we smooth the non-differentiable cut-off in the definition of ADFs via a Gaussian cumulative distribution function (CDF) centered at the cut-off (details on differentiable density, RDF and ADF in Supplementary Method 3). 

\subsection*{Stress-strain relations}
Computing the virial stress tensor $\bm{\sigma}^V$ for many-body potentials, e.g. NN potentials, under periodic boundary conditions requires special attention. This is due to the fact that most commonly used formulas are only valid for non-periodic boundary conditions or pairwise potentials \cite{Thompson2009}. Therefore, we resort to the formulation proposed by Chen et al. \cite{Chen2020}, which is well suited for vectorized computations in NN potentials.
\begin{equation}
    \bm{\sigma}^V = \frac{1}{\Omega} \left[ -\sum_{k=1}^{N_p} m_k \mathbf{v}_k \otimes \mathbf{v}_k - \mathbf{F}^T\mathbf{R} + \left( \frac{\partial U}{\partial \mathbf{h}}  \right)^T \mathbf{h} \right]  \ ,
\label{eq:stress_tensor}
\end{equation}
where $N_p$ is the number of particles, $\otimes$ represents the dyadic or outer product, $m_k$ and $\mathbf{v}_k$ are mass and thermal excitation velocity of particle $k$, $\mathbf{R}$ and $\mathbf{F}$ are $(N_\mathrm{p} \times 3)$ matrices containing all particle positions and corresponding forces, $\mathbf{h}$ is the $(3 \times 3)$ lattice tensor spanning the simulation box, and $\Omega = \det(\mathbf{h})$ is the box volume.

Due to the equivalence of the ensemble averaged virial stress tensor $\langle \bm{\sigma}^V \rangle$ and the Cauchy stress tensor $\bm{\sigma}$ \cite{Subramaniyan2008}, we can compute the elastic stiffness tensor from MD simulations and compare it to continuum mechanical experimental data (details in Supplementary Method 5). In the canonical ensemble, the isothermal elastic stiffness tensor $\mathbf{C}$ can be calculated at constant strain $\bm \epsilon$ via the stress fluctuation method \cite{VanWorkum2005}:
\begin{equation}
    C_{ijkl} =\frac{\partial \langle \sigma^V_{ij} \rangle}{\partial \epsilon_{kl}} = \langle C^B_{ijkl} \rangle - \Omega \beta \left(\langle \sigma^B_{ij} \sigma^B_{kl} \rangle - \langle \sigma^B_{ij} \rangle \langle \sigma^B_{kl} \rangle\right) + \frac{N_p}{\Omega \beta}\left( \delta_{ik} \delta_{jl} + 
\delta_{il} \delta_{jk} \right) \ ,
\label{eq:stress_fluctuation_method}
\end{equation}
with the Born contribution to the stress tensor $\sigma^B_{ij} = \frac{1}{\Omega}\frac{\partial U}{\partial \epsilon_{ij}}$, the Born contribution to the stiffness tensor $C^B_{ijkl} =
\frac{1}{\Omega}\frac{\partial^2 U}{\partial \epsilon_{ij} \partial \epsilon_{kl}}$ and Kronecker delta $\delta_{ij}$. 
Eq. \eqref{eq:stress_fluctuation_method} integrates well into DiffTRe by reweighting individual ensemble average terms (Eq. \eqref{eq:weights_definition}) and combining the reweighted averages afterwards.
Implementing the stress fluctuation method in differentiable MD simulations is straightforward: AD circumvents manual derivation of strain-derivatives, which is non-trivial for many-body potentials \cite{VanWorkum2006}. 

\subsection*{Statistical mechanics foundations}
Thermodynamic fluctuation formulas allow to compute the gradient $\frac{\partial L}{\partial \bm \theta}$ from ensemble averages \cite{DiPierro2013, Wang2013b, Wang2014a}. Specifically, considering a MSE loss for a single observable $O(U_{\bm \theta})$ in the canonical ensemble \cite{DiPierro2013},
\begin{equation}
    \frac{\partial L}{\partial \bm \theta} = 2 (\langle O(U_{\bm \theta}) \rangle - \tilde O) \left[
    \langle \frac{\partial O(U_{\bm \theta}) }{\partial \bm \theta} \rangle - \beta \left( \langle O(U_{\bm \theta}) \frac{\partial U_{\bm \theta}}{\partial \bm \theta} \rangle
    -\langle O(U_{\bm \theta}) \rangle \langle \frac{\partial U_{\bm \theta}}{\partial \bm \theta} \rangle \right) \right] \ .
    \label{eq:DiffTRe_gradient}
\end{equation}
It can be seen that the AD routine in DiffTRe estimates $\frac{\partial L}{\partial \bm \theta}$ by  approximating ensemble averages in Eq. \eqref{eq:DiffTRe_gradient} via reweighting averages (Derivation in Supplementary Method 5). 
End-to-end differentiation through the reweighting scheme simplifies optimization by combining obtained gradients from multiple observables.
This is particularly convenient for observables that are not merely averages of instantaneous quantities, e.g. the stiffness tensor $\mathbf{C}$ (Eq. \eqref{eq:stress_fluctuation_method}).

\subsection*{DimeNet++}
We employ a custom implementation of DimeNet++ \cite{Klicpera2020, Klicpera2020b} that fully integrates into Jax MD \cite{Schoenholz2020}. Our implementation takes advantage of neighbor lists for an efficient computation of the sparse atomic graph.
We select the same NN hyperparameters as in the original publication \cite{Klicpera2020b} except for the embedding sizes, which we reduced by factor 4. This modification allowed for a significant speed-up while retaining sufficient capacity for the problems considered in this work. For diamond, we have reduced the cut-off to 0.2 nm yielding an atomic graph, where each carbon atom is connected to its 4 covalently bonded neighbors.
A comprehensive list of employed DimeNet++ hyperparameters is provided in Supplementary Method 6.

\begin{sloppypar}
\section{Data availability}
Simulation setups and trained DimeNet++ models have been deposited in \href{https://github.com/tummfm/difftre}{https://github.com/tummfm/difftre}.
The data generated in this study are provided in the paper or in the Supplementary information file.

\section{Code availability}
The code for DiffTRe and its application to the three test cases is available at \href{https://github.com/tummfm/difftre}{https://github.com/tummfm/difftre} \cite{Thaler2021}.
\end{sloppypar}

\section{Acknowledgement}
This is a post-peer-review, pre-copyedit version of an article published in Nature Communications. The final authenticated version is available online at: \href{http://dx.doi.org/10.1038/s41467-021-27241-4}{http://dx.doi.org/10.1038/s41467-021-27241-4}

\section{Contributions}
S.T. conceptualized, implemented, and applied the DiffTRe method and conducted MD simulations as well as postprocessing.
S.T. and J.Z. planned the study, analyzed and interpreted results and wrote the paper.

\section{Competing interests}
The authors declare no competing interests.

\end{document}